\documentclass[preprint,superscriptaddress]{revtex4-1}

\usepackage{graphicx}
\usepackage{dcolumn}
\usepackage{bm}
\usepackage{amsmath}
\usepackage{caption}
\usepackage{setspace}
\usepackage[normalem]{ulem}
\usepackage[perpage]{footmisc}
\usepackage{lineno}

\begin{document}

\title{Predicting graphene's nonlinear-optical refractive response for propagating pulses}

\author{David Castell\'o-Lurbe}
\email{dcastell@b-phot.org}
\affiliation{Brussels Photonics, Department of Applied Physics and Photonics, Vrije Universiteit Brussel, Pleinlaan 2, 1050 Brussel, Belgium}
\author{Hugo Thienpont}
\affiliation{Brussels Photonics, Department of Applied Physics and Photonics, Vrije Universiteit Brussel, Pleinlaan 2, 1050 Brussel, Belgium}
\author{Nathalie Vermeulen} 
\affiliation{Brussels Photonics, Department of Applied Physics and Photonics, Vrije Universiteit Brussel, Pleinlaan 2, 1050 Brussel, Belgium}

\date{\today \rule[30pt]{0pt}{0pt}} 

\begin{abstract}
Nonlinear-optical refraction is typically described by means of perturbation theory near the material's equilibrium state. Graphene, however, can easily move far away from its equilibrium state upon optical pumping, yielding strong nonlinear responses that cannot be modeled as mere perturbations. So far, one is still lacking the required theoretical expressions to make predictions for these complex nonlinear effects and to account for their evolution in time and space. Here, this long-standing issue is solved by the derivation of population-recipe-based expressions for graphene's nonperturbative nonlinearities. The presented framework successfully predicts and explains the various nonlinearity magnitudes and signs observed for graphene over the past decade, while also being compatible with the nonlinear pulse propagation formalism commonly used for waveguides.
\end{abstract}

\maketitle

\section{Introduction}
Over the past decade there has been a rapidly growing interest in the theoretical \cite{Cheng1, Cheng2} and experimental \cite{Hendry, Chen, Ciesielski, Demetriou, Vermeulen1, Dremetsika, Takhur, Wang, Jiang, Alexander, Vermeulen2} investigation of nonlinear-optical refraction in graphene using free-space and waveguided excitation configurations. This new research field was pioneered by the experiments of Hendry \emph{et al.} showing an exceptionally large effective graphene nonlinearity $|\chi^{(3)}_\mathrm{gr}|\sim 10^{-7}\,\mathrm{esu}$ \cite{Hendry} or, equivalently, an effective nonlinear index $|n_{2,\mathrm{gr}}|\sim10^{-13}\,$m$^2$\,W$^{-1}$. The research results reported since then seem to point in different directions, and have made it a major challenge to fully understand graphene's nonlinear-optical behavior. The experimental data include both positive- \cite{Chen, Takhur} and negative-valued \cite{Demetriou, Vermeulen1, Dremetsika, Vermeulen2} effective nonlinearities with a magnitude compatible with that of Hendry's experiments \cite{Hendry}, as well as much smaller nonlinearities \cite{Wang}. From the theory point-of-view, calculations for the perturbative nonlinearity $\chi^{(3)}_\mathrm{gr}$ \cite{Cheng1, Cheng2} give rise to nonlinearities that are two orders of magnitude smaller than $|\chi^{(3)}_\mathrm{gr}|\sim 10^{-7}\, \mathrm{esu}$ measured in the aforementioned experiments. Strictly speaking,
using perturbation theory implies that the graphene
remains close to its initial equilibrium state, but this is not necessarily the case
\cite{Butcher}. 

Most of the experiments carried out so far were conducted using an exfoliated or chemical-vapor-deposition (CVD)-grown graphene sample without intentional doping. As such, when using optical excitation wavelengths, one-photon absorption (1PA) in the graphene layer is unavoidable, and the resulting free-carrier generation can give rise to nonlinearities \emph{outside} the perturbative regime with strong changes not only in temperature \cite{Jiang, Lui} but also in chemical potential \cite{Alexander}. These changes will be time- and space-dependent when dealing with propagating excitation pulses. What is more, our recent investigations on the evolution of the spectral bandwidth of pulses propagating in graphene-covered waveguides \cite{Vermeulen2} have shown the need for a nonperturbative treatment of graphene's nonlinear-optical refraction, with saturation playing an important role in the carrier dynamics. For telecom excitation wavelengths, we observed that this 'saturable photoexcited carrier refraction' yields a strong, negative nonlinearity from a phenomenological point of view, but we did not yet present a mathematical description for the refraction efficiency as a function of graphene's microscopic properties \cite{Vermeulen2}.

Microscopic models used for calculating the ultrafast carrier dynamics and the resulting nonlinear-optical response of graphene are generally built upon the equation of motion for the density matrix \cite{Butcher}. Over the years, these models have been applied for describing the nonlinear-optical physics in both large-area graphene and graphene nanoribbons \cite{Cheng1, Cheng2, Strouken, Malic, Marini, Baudisch}. Although many-particle interactions such as carrier-carrier and carrier-phonon scattering can be rigorously included in this framework \cite{Strouken, Malic}, the complexity and computational intensiveness of this modeling motivate simpler treatments of this microscopic physics based on the relaxation-time approximation. Indeed, this approach was employed \emph{e.g.} in \cite{Cheng2}, where graphene's perturbative optical nonlinearity was calculated solving the semiconductor Bloch equations, and also in \cite{Marini, Baudisch}, where nonperturbative phenomena in graphene were studied by means of the generalized Bloch equations for massless Dirac fermions \cite{Ishikawa}. Whereas the latter calculations are suitable for time-dependent excitations outside the perturbative regime, another approach is needed, however, to deal with propagating optical pulses \emph{e.g.} in waveguides, in which case the nonlinear response for a given electric field is typically described by means of a closed-form expression.  

In this letter, we show that graphene's nonlinear refraction efficiency for time- and space-dependent excitations outside the perturbative regime can be predicted using an easily accessible formalism based on the so-called population recipe \cite{Butcher}. More specifically, the refraction efficiency can be obtained from the excitation-induced instantaneous change in both temperature and chemical potential, and from the resulting instantaneous change in the \emph{linear} conductivity of graphene, in line with the population recipe. We derive an expression that links the conductivity change with the commonly used effective nonlinear index $n_{2,\mathrm{gr}}$ and with the free-carrier refraction (FCR) coefficient $\sigma_\mathrm{FCR}$ introduced in \cite{Vermeulen2}. Finally, we showcase the validity of our formalism for a wide variety of experiments in free space and waveguides.

\section{Rate equation and population recipe}

Semiconductor nonlinearities resulting from the optical excitation of free carriers can often be described by means of the so-called population recipe, where the equilibrium carrier concentrations that enter into the \emph{linear} conductivity are replaced by the photoexcited carrier concentrations \cite{Butcher}. In this section we investigate whether this general ansatz for semiconductors could also hold for graphene. Let us assume a single-mode electric field propagating in the $z$-direction, $\mathbf{E}(x, y, z, t) = A(z,t)/(\sqrt{4\,N})\,\mathbf{e}(x, y,\omega_0)\,e^{-i\omega_0 t}e^{i\beta(\omega_0)z} + \mathrm{c.c.}$, where $\omega_0$ is the mean spectral frequency of the light beam, $N$ is a normalization constant such that $|A(z,t)|^2$ equals the power, $\mathbf{e}$ denotes the transverse distribution of the electric field, and $\beta$ corresponds to the propagation constant. If a current density $\mathbf{J}$ is produced due to the interaction between the electric field and a material such as graphene, exhibiting 1PA characterized by its conductivity $\sigma_{ij}(\omega_0)$, then the electromagnetic energy converted into free carriers per unit volume and unit time is given by $\partial_t w = \mathbf{J}\cdot\mathbf{E} \approx |A(z,t)|^2\,e_i^*\sigma_{ij}e_j/(2N)$, where the last equation assumes a cycle averaging of the field \cite{Jackson}. In view of this equation, the dependence on space and time coordinates in $w$ can be separated as $w(x,y,z,t)=u(z,t)v(x,y)$. Since $u$ can show complex dependences on $A$ as several carrier-related processes may exist, effective models are often employed to deal with these phenomena \cite{Vermeulen2, Lin}. Along the same lines, we will study the temporal dynamics of the photoexcited carrier concentration based on the following rate equation for $u$:

\begin{equation}\label{rate}
\partial_t u = (1-\frac{u}{u_\mathrm{sat}})|A(z,t)|^2-\frac{u}{\tau_c},
\end{equation}

\noindent where $u_\mathrm{sat}$ and $\tau_c$ are phenomenological parameters that account for saturation and decay mechanisms, respectively. The 1PA-induced carrier concentrations evolving in $z$ and $t$ can then be computed by means of a spatial average (over $(x,y)$) of $w/(\hbar\omega_0)$, where $\hbar$ is the reduced Planck constant. In the Supporting Information (SI) we show that Equation~(\ref{rate}) can be derived from the generalized Bloch equations for massless Dirac fermions \cite{Ishikawa} on condition that the optical pulse width $T_0$ is sufficiently long ($T_0 \gtrsim 100\,$fs) and that the light intensity satisfies $I_0 \ll (1.8 \times 10^{-9}\,\mathrm{W}\,\mathrm{m}^2) / \lambda_0^4$\,\cite{Marini, Baudisch} with $\lambda_0 = 2\pi\,c/\omega_0$ and $c$ the speed of light in vacuum (\emph{e.g.} $I_0 \ll 3 \times 10^{14}\,$W/m$^2$ for $\lambda_0 = 1550\,$nm). The implications of these conditions are further discussed in Section 4.

Now we explore how $u$ behaves far from $\partial_t u = 0$ solving Equation  (\ref{rate}) by the integrating factor method,

\begin{align}\label{u}
\frac{u(z,t)}{u_\mathrm{sat}} & =
\int_{-\infty}^{\tau}\,\exp\bigg[-\int_{\tau'}^{\tau}
\bigg(\frac{T_0}{\tau_\mathrm{sat}}U(z,\tau'')+\frac{T_0}{\tau_\mathrm{c}}\bigg)\mathrm{d}\tau''\bigg]\nonumber\\
&\times\frac{T_0}{\tau_\mathrm{sat}}U(z,\tau')\mathrm{d}\tau'\approx
\frac{|A(z,t)|^2/P_\mathrm{sat}}{1+|A(z,t)|^2/P_\mathrm{sat}},
\end{align}

\noindent where $|A(z,t)|^2=P_0\,U(z,\tau=t/T_0)$, with $P_0$ being the input peak power. To facilitate interpretation, a saturation time $\tau_\mathrm{sat} = u_\mathrm{sat}/P_0$ and a saturation power $P_\mathrm{sat} = u_\mathrm{sat}/\tau_c$ have been defined. The last step in Equation  (\ref{u}) assumes $T_0\gg\tau_\mathrm{sat}$ and $T_0\sim\tau_c$ \cite{Vermeulen2}.

Important here is that Equation~(\ref{u}) can be recovered from the steady-state solution of Equation  (\ref{rate}) ($\partial_t u = 0$) if the time-independent steady-state power $|A(z)|^2$ is replaced by the instantaneous power $|A(z,t)|^2$. Hence the photoexcited carrier densities in graphene could behave as carrier densities in `instantaneous' steady states. As such, graphene's non-perturbative response could still be well described by means of the steady-state linear conductivity but with the carrier density at equilibrium replaced by the photoexcited densities calculated along Equation~(\ref{rate}). This carrier-induced instantaneous change in the linear conductivity then translates into the nonlinear response of the material.

\section{Nonlinear index and FCR coefficient}
To implement the approach outlined above for modeling graphene's nonlinear response for propagating pulses in the non-perturbative regime, we need to find explicit functional dependences of the chemical potential $\mu$ and temperature $T$ on the carrier concentration. To determine these, we consider that the probability of occupation of a state with energy $E$ within the time scale of the excitation pulse is given by instantaneous Fermi-Dirac distributions. The density of states in graphene near the Dirac point is approximately $\rho_\mathrm{gr} = 2|E|/(\pi(\hbar v_F)^2)$, where $v_F\approx 10^6\,$m\,s$^{-1}$ is the Fermi velocity. The 2D densities of electrons $n$ and holes $p$ are calculated as $n(\mu, T)=2/(\pi(\hbar v_F)^2)\,\mu^2\,J_1(\zeta)/\zeta^2$ and $p(\mu, T)=2/(\pi(\hbar v_F)^2)\,\mu^2\,J_1(-\zeta)/\zeta^2$, where $\zeta=\mu/(k_B T)$, $J_1(\zeta)$ is a Fermi-Dirac integral and $k_B$ is the Boltzmann constant \cite{Fang}. We aim at deriving closed expressions for $\mu=\mu(n)$ and $T=T(n)$ taking into account the electroneutrality condition, which relates $n$ and $p$. To illustrate our approach we assume an initial hole density $p_0$ \cite{Vermeulen1, Dremetsika, Vermeulen2}, so $p = p_0 + n$.

To obtain explicit solutions $\mu=\mu(n)$ and $T=T(n)$, we use two fitting functions as 
specified in the SI and after some algebra derive:

\begin{equation}\label{chemPot}
\mu = -\bigg(\frac{\pi}{2\kappa_\mu}\bigg)^{1/2}\hbar\,v_F\,
\frac{|\log(\eta)|}{\bigg(1+\displaystyle\frac{1}{2\kappa_\mu}\log^2(\eta)\bigg)^{1/2}}\,(n + p_0)^{1/2}
\end{equation}


\begin{equation}\label{temp}
k_B T = \bigg(\frac{\pi}{2\kappa_T}\bigg)^{1/2}\,\hbar\,v_F
\frac{(n + p_0)^{1/2}}{\bigg(1+\displaystyle\frac{1}{2\kappa_\mu}\log^2(\eta)\bigg)^{1/2}}
\end{equation}

\noindent with $\eta=n/(n+p_0)$, $\kappa_\mu = 2.682$ and $\kappa_T = 0.957$. Note that for $n=0$, $T=0$ and $\mu = -\hbar v_F(\pi p_0)^{1/2}$ are retrieved.

Now we turn to the imaginary part of graphene's conductivity at finite temperature, which is given by \cite{Cheng2}

\begin{equation}\label{sigmaT}
\mathrm{Im}\big[\sigma^{(1)}_{yy}(\omega, \mu, T)\big] = \frac{1}{2 k_B T}\int_{-\infty}^{+\infty}\frac{\mathrm{Im}\big[\sigma_{yy}^{(1)}(\omega, \tilde{x},0)\big]}{1 + \mathrm{cosh}\bigg(\displaystyle\frac{\tilde{x}-\mu}{k_B T}\bigg)}\mathrm{d}\tilde{x},
\end{equation}

\vspace{-\baselineskip}

\begin{align}\label{sigma0}
\mathrm{Im}\big[\sigma^{(1)}_{yy}(\omega, \mu, 0)\big] & =\frac{\sigma_0}{\pi}\bigg[\frac{4|\mu|\hbar\omega}{(\hbar\omega)^2 + \Gamma_\mathrm{intra}^2}\nonumber 
\\ & -\frac{1}{2}\log\bigg(\frac{(2|\mu|+\hbar\omega)^2 +\Gamma_\mathrm{inter}^2}{(2|\mu|-\hbar\omega)^2 + \Gamma_\mathrm{inter}^2}\bigg)\bigg].
\end{align}

\noindent Here $\sigma_0 = e^2/4\hbar$ is the universal conductivity, with $-e$ being the electron charge, and $\Gamma_\mathrm{intra}$ and $\Gamma_\mathrm{inter}$ are 
scattering parameters for intraband and interband transitions, respectively. This result accounts for the interband and intraband motion of electrons around the Dirac cone at the single-particle level \cite{Cheng1, Cheng2}. Equation~(\ref{sigmaT}) with $\mu$ and $T$ as defined in Equation  (\ref{chemPot}) and (\ref{temp}), incorporating the effect of photoexcited carriers $n[A(z,t)]$, accounts for the carrier-induced instantaneous change in the imaginary part of the conductivity $\mathrm{Im}\big[\Delta\sigma_{yy}^{(1)}\big] = \mathrm{Im}\big[\sigma^{(1)}_{yy}(n[A(z,t)]) - \sigma^{(1)}_{yy}(0)\big]$, and provides a closed-form expression for graphene's time- and space-dependent nonlinear refractive response in the non-perturbative regime.  

It should be noted that the quantity often measured in nonlinear experiments is the temporal average of the refractive index change $\int_{-\infty}^{+\infty}\Delta n(t)\,I(t)\,\mathrm{d}t/\int_{-\infty}^{+\infty}I(t)\,\mathrm{d}t$ where $I(t)$ represents the instantaneous light intensity, and subsequently, the nonlinear index defined via $n_{2}=\int_{-\infty}^{+\infty}\Delta n(t)\,I(t)\mathrm{d}t/\int_{-\infty}^{+\infty}I^2(t)\mathrm{d}t$. Modeling graphene as a thin layer of 3D material as done in many experiments \cite{Hendry, Chen, Ciesielski, Vermeulen1, Dremetsika, Takhur, Wang}, we derive the following expression to determine the magnitude and sign of graphene's $n_{2}$ (see SI):

\begin{equation}\label{n2eff}
n_{2,\mathrm{gr}} \sim -\frac{\pi\,\alpha^2}{4}
\frac{\lambda_0}{d_\mathrm{gr}}\frac{\mathrm{Im}(\Delta\sigma_{yy}^{(1)}(n_\mathrm{sat}/2)/\sigma_0)}{n_\mathrm{sat}/2}\frac{\mathrm{min}(\tau_c,T_\mathrm{FWHM})}{\hbar\omega_0},
\end{equation}

\noindent where $\alpha = e^2/(4\pi\varepsilon_0\hbar c)\approx 1/137$ denotes the fine-structure constant, $d_\mathrm{gr}=0.33$\,nm is the effective graphene thickness,  
$n_\mathrm{sat}$ is the saturation value of $n$, and $T_\mathrm{FWHM}$ indicates the full-width-at-half-maximum of the pulse duration. 

In contrast to the works reporting a nonlinear index for graphene, we already analyzed our experiments with graphene-covered waveguides in \cite{Vermeulen2} in terms of a nonlinear carrier-refraction response of the form $-\sigma_\mathrm{FCR}N_c$ embedded in the commonly used nonlinear pulse propagation formalism. Here  $N_c$ is a 1D carrier density defined at each $z$-distance along the waveguide. Using \cite{Lin} as our starting point we derive an expression for our FCR coefficient:

\begin{gather}\label{sigmaFCR}
\sigma_\mathrm{FCR} = \alpha\frac{v_F}{c}\frac{\mathrm{Im}(\Delta\sigma_{yy}^{(1)}(\tilde{N}_\mathrm{sat}/2)/\sigma_0)}{\tilde{N}_\mathrm{sat}/2}\frac{\lambda_0}{4\,n_\mathrm{eff}}\frac{\displaystyle\int_{w_\mathrm{graph}}|e_y(x_\mathrm{graph}, y)|^2\mathrm{d}y}{\displaystyle\int_S|e_y(\mathbf{x}_t)|^2\mathrm{d}S},
\end{gather}

\noindent where $e_y$ represents the $y-$component, parallel to the graphene sheet of width $w_\mathrm{graph}$, of the waveguide's quasi-transverse electric-field mode, $\tilde{N}_\mathrm{sat}=(v_F/\omega_0)n_\mathrm{sat}^{1/2}$, and $n_\mathrm{eff}$ denotes the waveguide effective index. We point out that $\sigma_\mathrm{FCR}$ does not depend on $d_\mathrm{gr}$ nor on the pulse duration, making it a robust measure for graphene's nonlinear response.

\begin{figure}[tb]
\includegraphics{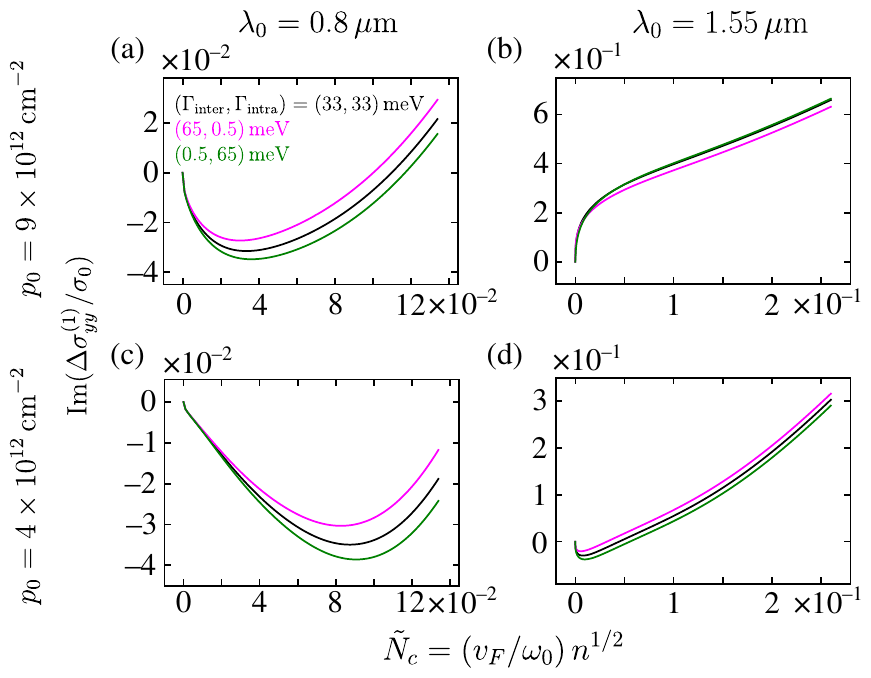}
\caption{\label{fig1} (a)-(d) Imaginary part of the calculated dynamic conductivity change of graphene under different conditions (see text and \cite{Cheng2}) as a function of the normalized square-root of the carrier density ranging up to $n_\mathrm{sat}=10^{17}$\,m$^{-2}$ \cite{Vermeulen2}.}
\end{figure}

\begin{table*}[t]
\begin{center}
\resizebox{\textwidth}{!}{
    \begin{tabular}{c c c c c c c c}
    \hline
    Work & $\lambda$ (nm) & $p_0^\mathrm{estimated}$ (cm$^{-2}$) & $T_\mathrm{FWHM}$ (ps) & 
    $I_0$ (W\,m$^{-2}$) & $n_{2,\mathrm{gr}}^\mathrm{exp}$ (m$^2$\,W$^{-1}$) &
    $\mathrm{Im}(\Delta\sigma_{yy}^{(1)}(n_\mathrm{sat}/2)/\sigma_0)$ & $n_{2,\mathrm{gr}}^\mathrm{theo}$ (m$^2$\,W$^{-1}$) \\ \hline
    
    \cite{Chen} & $733$ & $4\times10^{12}$ & $0.1$ & $9\times10^{14}$ & $1\times10^{-13}$ & $-3.6\times10^{-2}$ & $2\times10^{-14}$ \\
    
    \cite{Takhur} & $900$ & $4\times10^{12}$ & $0.1$ & $5\times10^{13}$ & $1\times10^{-12}$ & $-1.0\times10^{-2}$ & $1\times10^{-14}$ \\
    
    \cite{Takhur} & $900$ & $4\times10^{12}$ & $0.475$ & $5\times10^{13}$ & $2\times10^{-12}$ & $-1.0\times10^{-2}$ & $5\times10^{-14}$ \\
    
    \cite{Dremetsika} & $1550$ & $4\times10^{12}$ & $3.8$ & $5\times10^{12}$ & 
    $-2\times10^{-12}$ & $0.18$ & $-6\times10^{-12}$ \\
    
    \cite{Vermeulen1} & $1553$ & $4\times10^{12}$ & $3$ & $2\times10^{13}$ & $-4\times10^{-13}$ & $0.17$ & $-5\times10^{-12}$ \\
    
    \cite{Dremetsika} & $1600$ & $4\times10^{12}$ & $0.18$ & $5\times10^{12}$ & 
    $-1\times10^{-13}$ & $0.19$ & $-1\times10^{-12}$ \\
    
    \cite{Demetriou} & $2400$ & $4\times10^{12}$ & $0.1$ & $2\times10^{14}$ & $-3\times10^{-13}$ & $0.74$ & $-5\times10^{-12}$ \\ \hline
    
    \cite{Wang} & $355$ & $9\times10^{12}$ & $10$ & $1\times10^{14}$ & $5\times10^{-16}$ & $-1.4\times10^{-2}$ & $2\times10^{-14}$ \\
    
    \cite{Ciesielski} & $800$ & $9\times10^{12}$ & $0.015$ & $3\times10^{15}$ & 
    $\pm2\times10^{-11}$ & $-1.0\times10^{-2}$ & $1\times10^{-15*}$ \\
    
    \cite{Hendry} & $1000$ & $9\times10^{12}$ & $6$ & $1\times10^{13}$ & $\pm4\times10^{-13}$ & $6.1\times10^{-2}$ & $-8\times10^{-13}$ \\
    
    \hline\hline
    Work & $\lambda$ (nm) & $p_0^\mathrm{measured}$ (cm$^{-2}$) & $T_\mathrm{FWHM}$ (ps) & $I_0$ (W\,m$^{-2}$) & $\sigma_\mathrm{FCR}^\mathrm{exp}$ & $\mathrm{Im}(\Delta\sigma_{yy}^{(1)}(\tilde{N}_\mathrm{sat}/2)/\sigma_0)$ & $\sigma_\mathrm{FCR}^\mathrm{theo}$ \\ \hline
    
    \cite{Vermeulen2} & $1563$ & $6.5\times10^{12}$ & $3$ & $3\times10^{12}$ & $(1\pm0.2)\times10^{-5}$ &  $0.26$ & $0.7\times10^{-5}$ \\
    \hline
    
    \end{tabular}
    }

    \caption{\label{table1} Comparative table of the magnitudes and signs of graphene's experimental $n_{2,\mathrm{gr}}$ and $\sigma_\mathrm{FCR}$ values and the theoretical values calculated by means of Equation  (\ref{n2eff}) with estimated parameter values $n_\mathrm{sat}=10^{17}$\,m$^{-2}$ and $\tau_c = 1$\,ps \cite{Vermeulen2, Sun} and Equation~(\ref{sigmaFCR}). Works \cite{Wang, Ciesielski, Hendry} were carried out with exfoliated graphene, and all others with CVD graphene. The symbol~$\pm$ indicates that the sign of $n_{2,\mathrm{gr}}^\mathrm{exp}$ could not be determined in those experiments. $^*$See text for a discussion of this case.}
    
\end{center}
\end{table*}

\section{Comparison with experiments in literature}
Most of the experiments to measure graphene's nonlinearity at optical wavelengths have been performed using exfoliated or CVD graphene without intentional doping. As such, for most of those graphene samples the exact $p_0$ value has not been measured. However, typical ranges of unintentional doping values for exfoliated and CVD graphene can be found in \cite{Caridad} and \cite{Ciuk}, respectively. Based on these works we will consider an average doping level of $p_0 = 9\times 10^{12}$\,cm$^{-2}$ for exfoliated graphene and an average doping level of $p_0 = 4\times 10^{12}$\,cm$^{-2}$ for CVD graphene. To illustrate the behavior of $n_{2,\mathrm{gr}}$, Figures \ref{fig1}(a-d) show the calculated $\mathrm{Im}(\Delta\sigma_{yy}^{(1)}/\sigma_0)$ vs. $\tilde{N}_c =(v_F/\omega_0)n^{1/2}$ at two commonly used wavelengths $\lambda_0 = 0.8\,\mu$m and $\lambda_0 = 1.55\,\mu$m and at the doping values for exfoliated and CVD graphene. Note that these graphs are robust against changes in the scattering rates $\Gamma_\mathrm{intra, inter}$. According to these results and Equation  (\ref{n2eff}), $n_{2,\mathrm{gr}}$ is expected to switch sign between $0.8\,\mu$m and $1.55\,\mu$m with positive (negative) values at the shorter (longer) wavelengths (see SI for a discussion in terms of interband and intraband contributions). This prediction is in full agreement with the experiments reported so far (see Table \ref{table1}) and shows the tunability of graphene's nonlinearity sign in the near-infrared. Our theory also puts forward the so far undetermined (negative) sign of $n_{2,\mathrm{gr}}$ in Hendry's seminal experiments \cite{Hendry}.

Besides predicting correct signs for $n_{2,\mathrm{gr}}$, Equation  (\ref{n2eff}) also provides orders of magnitude compatible with most $n_{2,\mathrm{gr}}^\mathrm{exp}$ values in Table \ref{table1} (including the highest values), within its one-order-of-magnitude precision. This correspondence is remarkable as our calculations were carried out with \emph{estimated} parameter values $n_\mathrm{sat}=10^{17}$\,m$^{-2}$ and $\tau_c = 1$\,ps \cite{Vermeulen2, Sun} (see Table \ref{table1}). Equation~(\ref{n2eff}) also recovers the scaling of $n_{2,\mathrm{gr}}$ with pulse duration in the sub-ps excitation regime as  experimentally observed in \cite{Takhur, Dremetsika}. Furthermore, as discussed in the SI, according to our theory the nonlinear response is mainly introduced by the dynamical behavior of the temperature rather than chemical potential and its switching in sign is mainly produced by changes in the interband contribution.    

When verifying again the conditions defining the validity area of our theory (namely $T_0 \gtrsim 100\,$fs and $I_0 \ll (1.8 \times 10^{-9}\,\mathrm{W}\,\mathrm{m}^2) / \lambda_0^4$) we find that these conditions are essentially fulfilled for all works enlisted in Table \ref{table1} except for \cite{Ciesielski}, where $15$-fs-long pulses at high optical intensities were used. The latter explains why our theory underestimates $n_{2,\mathrm{gr}}^\mathrm{exp}$ reported in \cite{Ciesielski} by several orders of magnitude (see result marked with an asterisk in Table \ref{table1}). In experiments with very short pulses such as in \cite{Ciesielski}, the intraband carrier thermalization that follows right after the interband excitation might still be ongoing at the end of the pulse duration. Under such circumstances, the intraband contribution to the conductivity change might not be important, in which case a much higher positive nonlinearity value could be produced by theory (see SI). Further theoretical developments will be required to check this hypothesis. This being said, the accessible theory presented here does yield a good match with all other data in Table \ref{table1} observed for ps and sub-ps pulse lengths. This underlines the general applicability of our model for a wide range of excitation conditions.

Finally, we want to emphasize that $\sigma_\mathrm{FCR}$ as specified in Equation  (\ref{sigmaFCR}) is a more robust parameter than $n_{2,\mathrm{gr}}$ to quantify graphene's nonlinear response outside the perturbative regime. For the $\sigma_\mathrm{FCR}$ value
that we measured in graphene-covered waveguides \cite{Vermeulen2}, we find an excellent agreement with our theoretical prediction (see bottom of 
Table \ref{table1}).

In conclusion, with our population-recipe-based theory we can adequately predict graphene's nonlinear refractive response for propagating pulses outside the perturbative regime. With our expressions for $n_{2,\mathrm{gr}}$ and  $\sigma_\mathrm{FCR}$, one can readily calculate how the nonlinear effects evolve over time and distance. Our work explains the differences in sign and magnitude of graphene's nonlinear coefficients measured over the past decade, and also provides the long-needed framework for conceptualizing, designing, 
and optimizing graphene-enhanced nonlinear-optical (on-chip) devices.

\section*{Funding Information}
Support from ERC-FP7/2007-2013 grant 336940 and the Research Foundation Flanders (FWO) under Grants G005420N, G0F6218N (EOS-convention 30467715), VUB-OZR and FWO postdoctoral fellowship (147788/12ZN720N).

%

\end{document}